\documentclass[10pt,a4paper,twocolumn]{article}
\usepackage[margin=1in]{geometry}
\usepackage[utf8x]{inputenc}  
\usepackage{amsmath}
\usepackage{graphicx}
\graphicspath{ {./} }

\usepackage[super, sort&compress]{natbib}
\usepackage{natmove}

\usepackage{caption}
\usepackage{subcaption}
\usepackage{siunitx}
\usepackage[version=4]{mhchem}

\usepackage{varioref}
\usepackage[capitalize]{cleveref}

\newcommand{\boldq}{\ensuremath{\mathbf{q}}}
\newcommand{\qpt}{\boldq{}-point}
\newcommand{\codename}[1]{{\sc #1}} 

\title{A fast approximate method for variable-width broadening of spectra}
\author{Jessica Farmer and Adam J. Jackson\\
        Scientific Computing Department, Science and Technology Facilities Council,\\
        Rutherford Appleton Laboratory, Oxfordshire, UK}
\date{January 2022}

\begin{document}

\maketitle

\section{Abstract}
Spectral data is routinely broadened in order to improve appearance, approximate a higher sampling level or model experimental measurement effects.
While there has been extensive work in the signal processing field to develop efficient methods for the application of fixed-width broadening functions, these are not suitable for all scientific applications ---
for example, the instrumental resolution of inelastic neutron scattering measurements varies along the energy-transfer axis.
Na{\"i}ve application of a kernel to every point has $O(N \times M)$ complexity and scales poorly for a high-resolution spectrum over many data points.
Here we present an approximate method with complexity $O(N + W\times M \log M)$, where $W$ scales with the \emph{range} of required broadening widths;
in practice the number and cost of mathematical operations is drastically reduced to $N$ polynomial evaluations and a modest number of discrete Fourier transforms.
Applications are demonstrated for Gaussian interpolation of density-of-states data and to instrumental resolution functions.
We anticipate that these performance improvements will assist application of resolution functions inside fitting procedures and interactive tools.

\section{Introduction}
Many scientific properties are \emph{spectra}, i.e. continuous functions.
Experimental measurements tend to involve a degree of sampling (e.g. of photons in an optical measurement) and discretisation (e.g. finite detector sizes, measurement timesteps),
from which the ``true" underlying spectrum is approximated.
Computational simulations may have access to high precision floating-point operations
but still involve sampling in the form of numerical integration and finite-different approximations.
Such data is then routinely broadened in order to improve appearance, approximate a higher sampling level or model experimental measurement effects.\cite{o'haver}

A simple and widespread form of broadening is convolution of a spectrum $f(x)$ with a Gaussian function $g(x)$
\begin{equation}\label{convolve}
    (f * g)(x) = \frac{1}{\sqrt{2\pi}\sigma} \int e^{-\frac{x^2}{2\sigma^2}}f(x) dx,
\end{equation}
where $\sigma$ is the standard deviation of the Gaussian function.
This normal distribution may appropriately represent the resolution limits of a scientific instrument;
incorporating such quirks into simulated spectra allows for more direct comparison with real data.
This approach is generally simpler and more reliable than trying to correct the real data for the resolution of the instrument (deconvolution).\cite{fultz_2020}

Convolutional broadening is a well-explored topic in the fields of signal-processing and computer graphics.
Implementations using discrete Fourier transforms are available in popular libraries such as \codename{SciPy},
while recent innovations include applications of recursive filters and discrete cosine transforms.\cite{virtanen_scipy_2020,deriche_rachid_recursively_1993,sugimoto_efficient_2015}

Energy broadening is also routinely used in the computation of density-of-states (DOS) spectra, to suppress jagged sampling artefacts while preserving detail.
In both electronic structure calculations and phonon spectra, an energy distribution of states is sampled over a mesh of points in reciprocal (\boldq{}-) space.
If broadened using \cref{convolve}, a $\sigma{}$ value that gives smooth lines in the expected places may also be too broad for sharp peaks;
ideally the Gaussian width for each q-point should be proportional to the gradient of the energy with respect to \boldq{}.\cite{PhysRevB.75.195121}


Instrumental broadening can have multiple origins including incident pulse profiles, beam divergence/reflections and detector dimensions;
in the case of inelastic neutron scattering (INS) spectrometers a transformation from time-of-flight to energy-transfer domains can give a strong energy dependence.
These can be modelled to give a function of energy (e.g. for the TOSCA or VISION indirect-geometry neutron spectrometers);\cite{mitchell_vibrational_2005,seegerResolutionVISIONCrystalanalyzer2009}
in more complex cases the function may need to be recalculated to account for variable incident energy
or emerge from stochastic simulations using a code such as \codename{McStas} or \codename{Horace}.\cite{willendrupMcStasIntroductionUse2020,ewingsHoraceSoftwareAnalysis2016}
While broadening of a statistical or thermal origin tends to be Gaussian, optical lifetime broadening is Lorentzian in character
and varies between the individual excitations making up the spectrum;
X-ray absorption spectroscopy has been modelled with width parameters depending on both the initial core state and final energy,
while Raman spectroscopy has been simulated with temperature-dependent phonon lifetimes from three-phonon calculations.\cite{muller_x-ray_1982,skelton_vibrational_2015}

In these cases, broadening with a variable-width kernel can be implemented by broadening each sample individually and binning/summing to form the broadened spectrum. However, an exact implementation of this is computationally expensive: for input data of length $N$, there are $N$ evaluations of the kernel on a grid of length $M$, which are then summed. Therefore, exact variable-width broadening has $O(N \times M)$ complexity, which for large $N$ and $M$ will become slow.
Performance will be of particular concern when broadening inside the main loop of a fitting procedure or interactive visualisation.

A number of existing software packages implement variable-width broadening, for example \codename{OptaDOS} \cite{morris_optados_2014}, \codename{PyAstronomy}  \cite{pya}, \codename{Eniric} \cite{Neal2019} and a python module \codename{varconvolve} \cite{varconvolve}.
Not all of these are designed to apply broadening to INS spectra, but the underlying methods are the same. For \codename{OptaDOS} and \codename{PyAstronomy}, exact implementations of variable width broadening are available; if users want a faster option, the \codename{PyAstronomy} documentation suggests using fixed width broadening. The \codename{Eniric} code offers rotational broadening with a variable-width kernel, distributing the convolution of each point across parallel tasks.
The python package \codename{varconvolve} offers a different approach to variable-width convolution. Rather than change the width of the kernel with position along the x-axis, the input spectrum is instead warped to achieve the variability in the kernel. The warped spectrum can then be convolved with a fixed width kernel before the spectrum is then unwarped to its original scale. This technique reduces the number of Gaussian kernel evaluations to $M$, as only one fixed width convolution is carried out. However, warping the spectrum by interpolation introduces loss of accuracy. Another drawback of this method is that the broadening widths must be defined as a smooth function of $x$.
Whilst this can be true for instrumental broadening, in adaptive DOS broadening each datapoint is treated individually and peaks in the same energy bin can have different widths. In INS simulation code \codename{AbINS} \cite{dymkowski_abins_2018} (distributed as part of \codename{Mantid} \cite{arnold_mantiddata_2014,noauthor_mantid_2013}) implements a fast variable width broadening method with similar drawbacks to the \codename{varconvolve} method. Here, for a small number of broadening widths, the entire spectrum is broadened using fixed-width convolution. Nearby spectra can then be interpolated using predetermined mixing weights to produce a spectrum broadened with a variable-width function.

Here we build on the approximate method for variable-width Gaussian broadening implemented in \codename{AbINS} and present an improved, generalised method, limiting the number of expensive kernel evaluations with minimal loss in accuracy.
The method is outlined formally and scientific applications are demonstrated.

\section{Method} \label{method}
\emph{
For simplicity, the method is formally outlined in this section for Gaussian broadening; however it is easily adapted to other smooth broadening functions such as the Lorentzian and we include some data for Lorentzian functions.
}

Rather than explicitly evaluate a Gaussian function for each data point, a small set of spectra are convolved with fixed-width Gaussian kernels using Fast Fourier Transforms (FFTs).
These spectra consist of weighted points that, when summed together, form suitable approximations to the desired Gaussian widths and intensities.

\subsection{Gaussian Approximation} \label{gaussian}
\begin{figure}
    \centering
    \begin{subfigure}[b]{\columnwidth}
        \centering
        \includegraphics[width=0.9\columnwidth, trim={0 1cm 0 0}, clip]{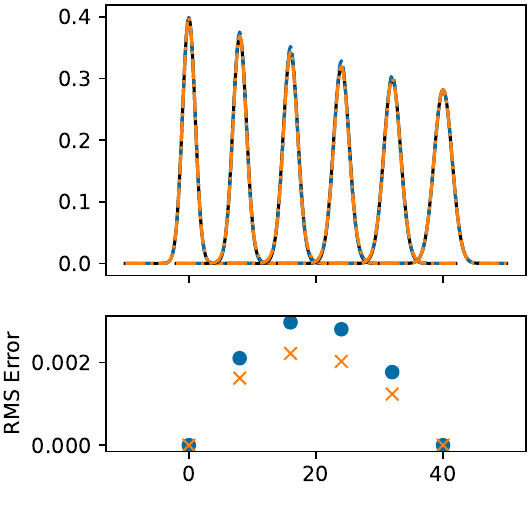}
        \caption{width factor $\alpha=\sqrt{2}$}
        \label{wfr2}
    \end{subfigure}
    
    \begin{subfigure}[b]{\columnwidth}
        \centering
        \includegraphics[width=0.9\columnwidth, trim={0 1cm 0 0}, clip]{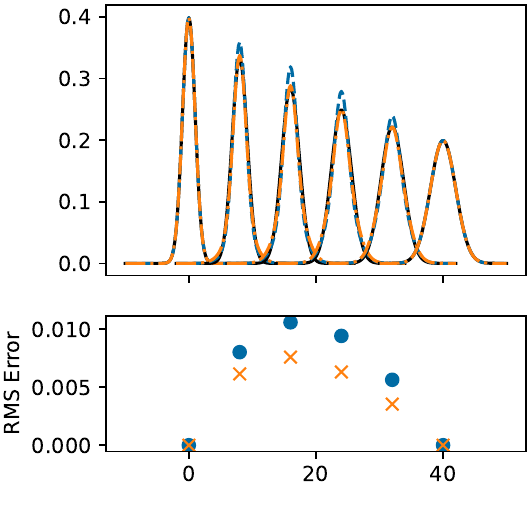}
        \caption{width factor $\alpha=2$}
        \label{wf2}
    \end{subfigure}
    
    \begin{subfigure}[b]{\columnwidth}
        \centering
        \includegraphics[width=0.9\columnwidth]{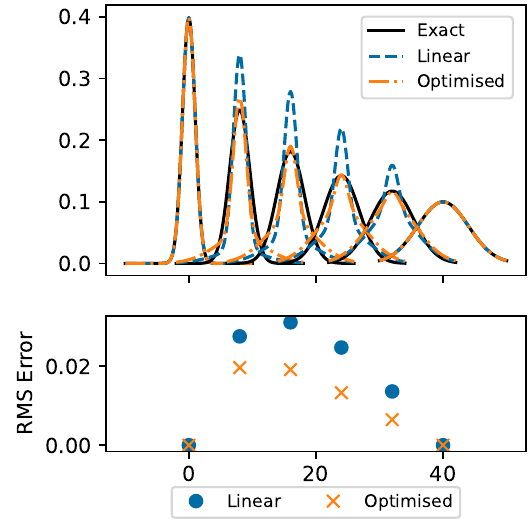}
        \caption{width factor $\alpha=4$}
        \label{wf4}   
    \end{subfigure}
    \caption{Gaussian functions $G_\alpha$ approximated as a mixture of end members $(G_1, G_2)$ over a range of width factors $\alpha=\frac{\sigma_{G_2}}{\sigma_{G_1}}$.
    Exact function is compared with na\"{i}ve linear interpolation and least-squares optimised fits. }
    \label{interp_compare}
\end{figure}

A small number of Gaussian kernels are explicitly calculated, covering the required range of $\sigma$ values for the variable-width broadening operation.
Broadening at intermediate $\sigma$ values can be approximated by a linear combination of the closest two exact Gaussians, one of which is narrower ($G_1$) and one wider ($G_2$),
\begin{equation}
    G_{approx} = (1-w)G_1 + wG_2, \label{eq:approximate-gaussian}
\end{equation}
where $w$ is the linear combination weight. The simplest way to choose $w$ is by linear interpolation: $w=(\sigma-\sigma_1)/(\sigma_2-\sigma_1)$,
but a better choice can be found by least-squares optimisation, as illustrated in \vref{interp_compare}.

A mixture of two Gaussian functions with the same mean but different $\sigma$ produces a leptokurtic distribution, which compared to a Gaussian distribution has a more prominent peak and longer tails.
A distribution is leptokurtic if its excess kurtosis $\kappa$ is positive:
\begin{equation}
    \kappa = \frac{\mu_4}{\mu_2^2} -3,
\end{equation}
where $\mu_i$ denotes the $i$th central moment of the distribution. Kurtosis can be used as a measure of the quality of a Gaussian approximation \cite{refId0}. For a mixture of two Gaussians, one with $\sigma=1$ and linear combination weight $w$, and the other with $\sigma$ equal to the width factor $\alpha$ and linear combination weight $(1-w)$, kurtosis is equal to
\begin{equation}
    \kappa = 3\frac{w + (1-w)\alpha^4}{(w + (1-w)\alpha^2)^2} - 3.
\end{equation}
When $0<w<1$ and $\alpha>1$, the kurtosis value decreases as $\alpha$ approaches 1, and therefore the linear combination becomes closer to a Gaussian distribution. Therefore, as illustrated in \cref{interp_compare}, a smaller width factor will yield more accurate approximations.

The choice of $\alpha$ also determines the number of Gaussians which will need to be explicitly calculated. The values of $\sigma$ for the exact Gaussians are chosen to be logarithmically spaced across the range of $\sigma$ values for the spectrum being broadened.
By using a common ratio between successive $\sigma$ samples,
we ensure that the relationship between $\sigma$ and the linear combination weights will be the same for each neighboring pair of Gaussians.
This relationship is determined by finding optimal linear combination weights for a small number of $\sigma$ values in the range $\sigma=1$ to $\sigma=\alpha$.
A polynomial is then be fitted to the computed weights to obtain the function $w(\sigma / \sigma_1)$,
allowing weights to be determined for any $\sigma$ value in that range.

\begin{figure}
    \centering
    \includegraphics[width=\columnwidth]{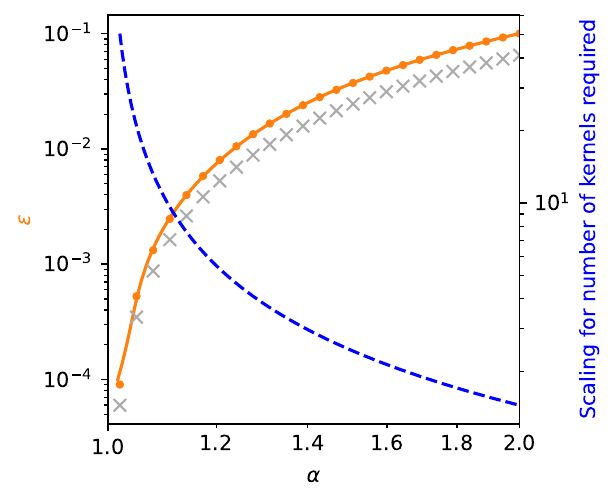}
    \caption{Relationship between interpolation error
    ($\epsilon$) and the width factor between exact Gaussian kernels ($\alpha$). 
    Orange dots show maximum error and green crosses show mean error for various width factors,
    while orange solid line shows maximum error estimate from \cref{spacing_eqn}.
    The relationship between width ratio and number of kernels spanning a fixed range of required $\sigma$ is indicated by blue dashed line.
    These error calculations use ideal mixing parameters over 100 points from -10 to 10.}
    \label{error_vs_efficiency}
\end{figure}

The logarithmic spacing means that the required number of broadening kernels
\begin{equation} \label{n_kernels}
    n = \lceil\frac{\log(\frac{\sigma_{max}}{\sigma_{min}})}{\log(\alpha)}\rceil + 1.
\end{equation}
Gaussians are explicitly calculated on the output data bins for widths in the set
\begin{equation} \label{sigma_samples}
    s = \{\, \alpha^{i} \times \sigma_{min} \mid i = 0,...,n\,\}.
\end{equation}
As $n$ increases, the computation time required to calculate the exact kernels increases: the choice of width factor is a trade-off between accuracy and efficiency, which should consider the application and the importance placed on highly accurate results.
This is illustrated in \cref{error_vs_efficiency}, showing the maximum area difference between the actual and estimated Gaussian for width factors $\alpha$ ranging from \numrange{1.02}{2}. 
A polynomial can be fitted to this data to obtain a fast estimate of appropriate spacing for a given error tolerance. A range of fits are illustrated in the Supplementary Information: the authors suggest
\begin{align} \label{spacing_eqn}
\alpha &= 6.66711679 + 9.65316879 x + 7.59554302 x^2 \nonumber \\
       &\quad + 3.34967211 x^3 + 0.845072779 x^4 \nonumber \\
       &\quad + 0.113549039 x^5 + 0.00628611391 x^5, \nonumber \\
       \intertext{where}
       -4 &< (x = \log_{10}\sigma) < -1.
\end{align}
for Gaussian broadening, and
\begin{align} \label{spacing_eqn_lorentz}
\alpha &= 10.4916374 + 14.9743160 x + 10.5868304 x^2 \nonumber \\
       &\quad + 4.13975214 x^3 + 0.923270724 x^4 \nonumber \\
       &\quad + 0.109892384 x^5 + 0.00541175294 x^5, \nonumber \\
       \intertext{where}
       -4 &< (x = \log_{10}\gamma) < -1.3.
\end{align}
for Lorentzian broadening,
where $\epsilon$ is the desired maximum error in the Gaussian approximations.
These polynomials were obtained by Chebyshev regression, which tends to minimise the maximum error along the curve.

\subsection{Convolution} \label{conv}
The convolution of two discrete data sets, $f$ and $g$, both of length $N$
\begin{equation} \label{convolution}
    (f * g)(x) = \sum_{k=0}^{N-1} f(k)g(x-k).
\end{equation}
Using the linear properties of convolution, we can avoid a sum over individual peaks. Suppose fixed-width convolution is carried out with the approximated Gaussian of \cref{eq:approximate-gaussian}. Then,

\begin{align}
f*G_{approx} &= f*((1-w)G_1 + wG_2) \\
             &= f*(1-w)G_1 + f*wG_2 \intertext{by distributivity}
                     &= (1-w)f*G_1 + wf*G_2
\end{align}
by associativity with scalar multiplication.

\begin{figure*}
    \centering
    \includegraphics[width=0.8\textwidth]{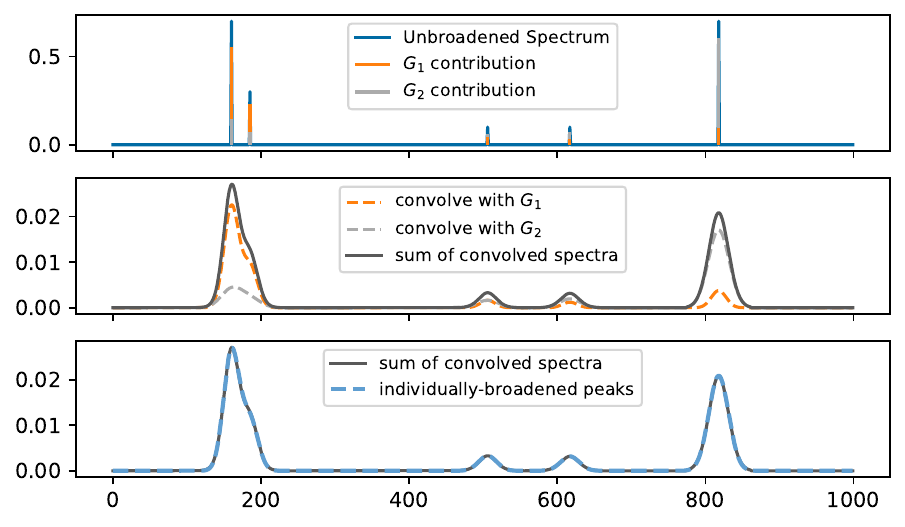}
    \caption{Schematic of the proposed approximate variable width broadening method. The top plot shows both the total unbroadened spectrum and its distribution into contributions to be treated with the two exact Gaussian kernels ($G_1$ and $G_2$). The middle plot shows the convolution of these contributions with $G_1$ and $G_2$ respectively, and the sum of the resulting spectra. Finally, the bottom plot compares this sum with a broadened spectrum produced by summation of individually-broadened peaks.
    In this example, $\sigma$ increases linearly along the x-axis, from 1 to $\sqrt{2}$. The standard deviations of $G_1$ and $G_2$ are 1 and $\sqrt{2}$ respectively.}
    \label{illustrate_method}
\end{figure*}

Now rather than convolving once with an approximated Gaussian, two convolutions are performed with exact Gaussians, while the spectrum $f$ is multiplied by the linear combination weights $w$. If the points of the unbroadened spectrum are scaled by corresponding arrays of weights $(1-w)$ and $w$, we obtain two spectra of contributions from $G_1$ and $G_2$ respectively. These can be broadened with their corresponding (fixed-width) kernels by convolution and summed to give the overall broadened spectrum.
The approach is illustrated in \cref{illustrate_method}. In this example, broadening width increases linearly along the x-axis, from 1 to $\sqrt{2}$ and two exact Gaussians $G_1$ and $G_2$ have widths of 1 and $\sqrt{2}$ respectively. \\

In practice, the method is applied to much denser data than in \cref{illustrate_method}. INS spectra can have many thousands of individual peaks with broadening widths spanning multiple orders of magnitude.
Prior to the convolution step, the spectrum is binned onto a regular grid, typically of a few thousand bins.
In such cases FFT convolution is usually more efficient than direct calculation. Direct implementation of \cref{convolution} requires calculating $N$ multiplications and $N-1$ additions, leading to a complexity of $O(N^2)$.
FFT convolution uses the convolution theorem:
\begin{align}
\mathcal{F}(f * g) &= \mathcal{F}(f) \mathcal{F}(g) \\
f * g &= \mathcal{F}^{-1} \left[ \mathcal{F}(f) * \mathcal{F}(g) \right]
\end{align}
where $\mathcal{F}$ and $\mathcal{F}^{-1}$ are the Fourier transform and its inverse.
For data on a regular grid, FFT algorithms with computational complexity of $O(N\log N)$ are used and so FFT convolution has overall complexity $O(N\log N)$.

\subsection{Implementation Steps}
Using the concepts described in \cref{gaussian} and \cref{conv}, the approximate variable-width broadening method comprises the following steps:
\begin{enumerate}
    \item Compute the required width factor $\alpha$ with \cref{spacing_eqn} to reach the desired maximum error level.
    \item Determine the number of Gaussians to be exactly computed, and the values of $\sigma$ for these Gaussians using \cref{n_kernels} and \cref{sigma_samples} respectively.
    \item On a grid with evenly spaced bins, evaluate the Gaussians for the selected $\sigma$ values.
    \item Obtain a model function $w(\sigma)$ that provides optimal mixing weights for a given target function width
    \item Use this model to distribute the input data to a set of spectra (one per exact Gaussian)
    \item Convolve each spectrum with its respective kernel
    \item Sum together all convolved spectra
    
\end{enumerate}

\subsection{Performance and Accuracy}
Conventional variable-width broadening which involves the summation of N individual peaks on a grid of length M has a complexity of $O(N \times M)$. The fast, approximate method significantly reduces the number of Gaussians that have to be computed exactly, as shown in \cref{n_kernels}. As a result, the complexity of approximate variable-width broadening is $O(N + W \times M\log M)$, where N is the length of the input spectrum, $W = \log( \frac{\sigma_{max}}{\sigma_{min}})$, and M is the length of the regular grid on which the broadened spectrum is calculated. The first term, N, arises from the number of data points, which are looped over in order to apply the function $w(\sigma)$ (obtaining mixing weights) and bin the data.
The second term relates to the Gaussian evaluations, the number of which is governed by the $\sigma$ range for the spectrum being broadened, and the FFT convolution step at each width on a regular grid of length M.
It is the decoupling of N and the number of Gaussian evaluations which provides the reduced complexity of the approximate variable-width broadening method. With this approach, the complexity scaling is either driven by M or by N (if this is large enough relative to M), compared to N $\times$ M for the summation over peaks method.

\begin{figure}
    \centering
    \includegraphics[width=\columnwidth]{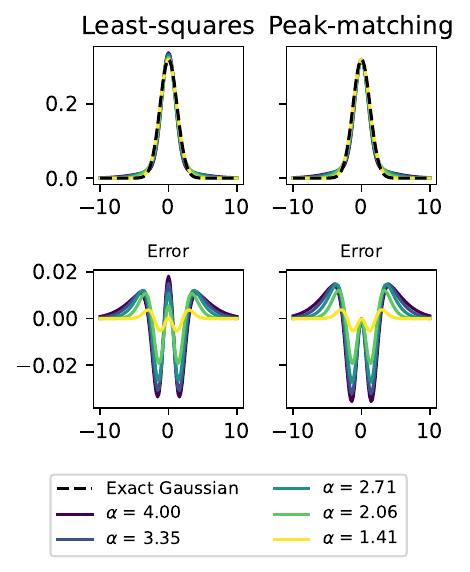}
    \caption{Variation in overall accuracy of Gaussian approximation with width factor.
    Upper plots show approximations to Gaussian with $\sigma=1.25$ produced using a linear combination of two Gaussians with $\sigma$ spaced by the specified width factor. Lower plots are obtained by subtracting exact Gaussian from approximations.
    The left column uses least-squares optimisation over whole data range to determine linear combination weights, whereas the on the right-hand-side the mixing parameters are chosen to match the exact peak height at $x = 0$.
    }
    \label{error_dist}
\end{figure}

The reduced computation time associated with the approximate method has the consequence of reduced accuracy in the broadening process. As shown in \cref{interp_compare}, approximating Gaussians using a linear combination of two exact Gaussian functions is not exact. It can be seen that even for a small width factor of $\sqrt{2}$, there is still a small amount of error in the approximations particularly at the peak of the Gaussian. Meanwhile for a wide width factor choice of 4, the approximated Gaussians have a clear leptokurtic distribution, losing the characteristics of a true Gaussian. \cref{error_dist} further illustrates how the approximation accuracy varies with the choice of width factor, showing how error is distributed across the Gaussian approximation. The left side of \cref{error_dist} shows the distribution of error when linear combination weights are determined by least-squares optimised interpolation, where error is present both at the peaks and the tails of the Gaussian.
(Equivalent plots are provided in SI for Lorentzian broadening.)

Depending on the application of the method, it may be preferable for this error to be distributed differently.
For example, if correct peak intensity is important then the error could be minimised at this point. The right-hand side of \cref{error_dist} shows the approximations produced when linearly interpolating the function value at $x=0$.
This forces the curve to pass through the peak of the true Gaussian while maintaining a constant area.
It leads to some redistribution of error into the tail region, compared with least-squares fitting, but may give a more visually satisfying result.

Another source of error comes from the requirement that data is binned onto a regular grid: if the choice of bin width is too coarse then fine detail in the data will be lost. This also applies to the Gaussian function evaluations which are evaluated on the same grid: if $\sigma$ is too small in relation to bin width, then the Gaussian will only be represented by a single point.
Techniques such as linear binning could be used to maintain accuracy with a slightly larger bin size, but this is not implemented here;
they are particularly powerful when working with higher-dimensional data.\cite{hallAccuracyBinnedKernel1996,wandFastComputationMultivariate1994}

\section{Application} \label{applications}
\subsection{Adaptive broadening of phonon DOS}
Adaptive broadening allows density-of-states (DOS) plots to be computed with both fine detail and smooth tails from a limited number of q-point samples.
This has been implemented in the \codename{OptaDOS} code and is routinely used for electronic structure DOS plots.\cite{morris_optados_2014} The python library \codename{Euphonic} \cite{fairEuphonicInelasticNeutron2022a,Fair_Euphonic_2023} implements adaptive broadening of the phonon DOS using the same method as \codename{OptaDOS}. Within \codename{Euphonic} the adaptive broadening step was found to be slow, overwhelming the performance benefit of reduced DOS sampling.

\begin{figure}
    \centering
    \begin{subfigure}[b]{\columnwidth}
        \includegraphics[width=\columnwidth]{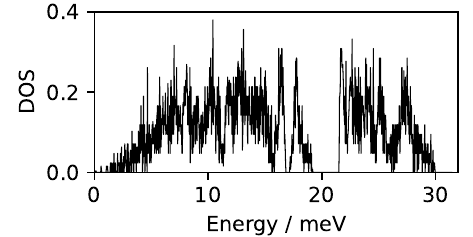}
        \caption{
        Unbroadened phonon DOS
        \label{dos_raw}}
    \end{subfigure}
        \begin{subfigure}[b]{\columnwidth}
        \includegraphics[width=\columnwidth]{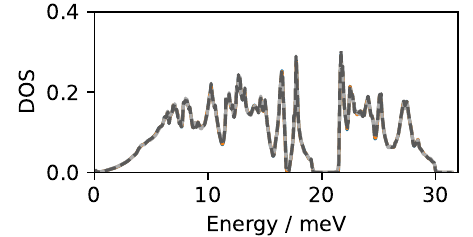}
        \caption{
        Phonon DOS with adaptive broadening
        \label{dos_example_broadened}
        }
    \end{subfigure}
    \begin{subfigure}[b]{\columnwidth}
        \includegraphics[width=\columnwidth]{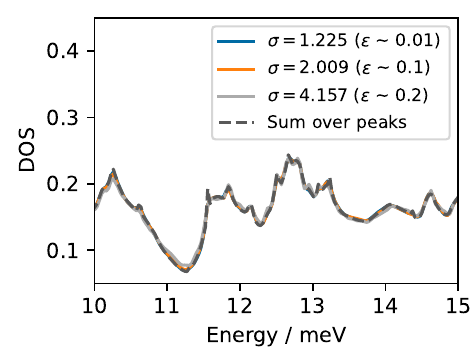}
        \caption{Magnified view of approximation artefacts
                \label{zoomed_example}}
    \end{subfigure}
    \caption{Application of approximate variable-width method to adaptive broadening of vibrational data.
A phonon DOS is computed using Euphonic with force constant data from the Kyoto Phonon Database
(elemental selenium, item \emph{mp-147-20180417})\cite{togoPhononDatabaseKyoto2015,jain_commentary_2013}.
For a variety of nominal error levels ($\epsilon = (0.01, 0.1, 0.2)$), corresponding $\sigma$ values (1.225, 2.009, 4.157) were estimated and used for approximate implementation of adaptive broadening.
(A Gaussian width is related to the local gradient of phonon mode in energy--momentum space).
Exact variable-width broadening of the spectrum by summation over peaks is overlaid as dashed line.
}
    \label{dos_example}
\end{figure}

The fast, approximate method reported here has been implemented in \codename{Euphonic}. When calling broadening functions, users are able to define the desired maximum error level in the Gaussian approximations, which determines the width factor to be used in the algorithm, calculated using \cref{spacing_eqn}. If the user does not specify the error limit, then a default of 0.01 is used. \cref{dos_example} shows a DOS that has been adaptively broadened, using both the summation over peaks method and the fast, approximate method with a selection of error levels.
The raw data sampling 18 bands over 1331 qpts is difficult to interpret, but a fixed wide broadening kernel would destroy sharp features.
It is visually clear that the fast, approximate method produces a very similar adaptively-broadened DOS to the exact summation over peaks, even with a large width factor with maximum error of 0.2 in the approximated Gaussians.
Inspecting a narrow region of the data in \cref{zoomed_example} there are some artefacts at this level, while a nominal error of 0.1 appears to closely follow the exact results.

For this particular dataset, the Gaussian $\sigma$ values ranged from the bin width \SI{0.032}{\milli\electronvolt} to \SI{0.759}{\milli\electronvolt} -- smaller $\sigma$ values obtained from the data gradients were rounded up to the bin width.
The exact sum over peaks required \num{23958} Gaussian kernels, whereas with a width factor 1.22 (nominal error 0.01) a total of 17 Gaussian functions are evaluated --- a reduction of 3 orders of magnitude.

\subsection{Instrumental resolution functions}
The resolution of time-of-flight INS instruments is limited by the size of detectors and width of neutron pulses.
This can lead to energy and \boldq{}-dependent broadening of the underlying scattering function, 
which may be applied to simulated spectra to facilitate analysis of experimental results.

\begin{figure*}
   \centering
    \begin{subfigure}[b]{\columnwidth}
        \includegraphics[width=0.9\columnwidth,trim={0.25cm 0.25cm 0.25cm 0.25cm},clip]{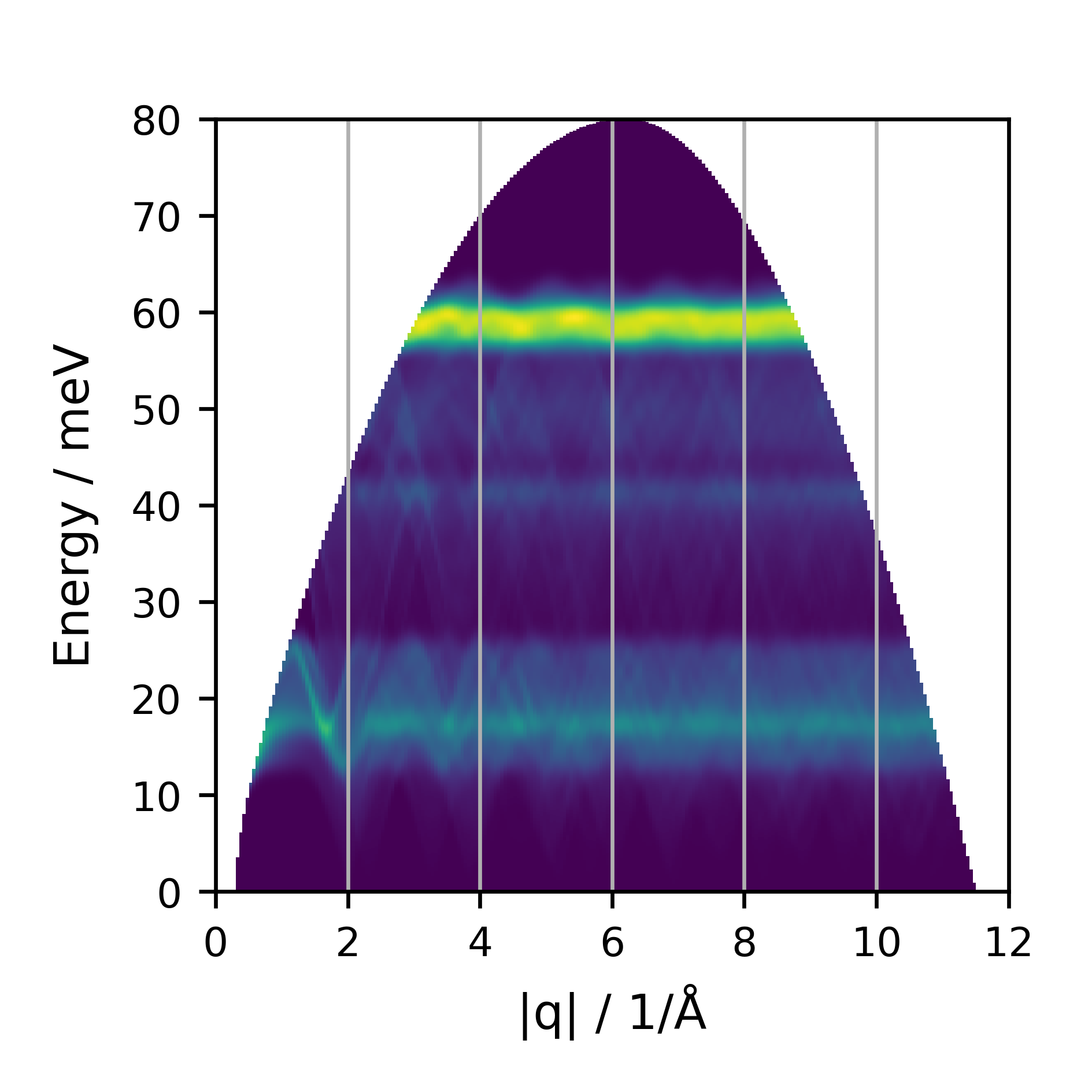}
        \caption{
        Fixed-width broadening: \SI{2}{\milli \electronvolt}
        \label{fig:merlin-2}}
    \end{subfigure}
        \begin{subfigure}[b]{\columnwidth}
        \includegraphics[width=0.9\columnwidth,trim={0.25cm 0.25cm 0.25cm 0.25cm},clip]{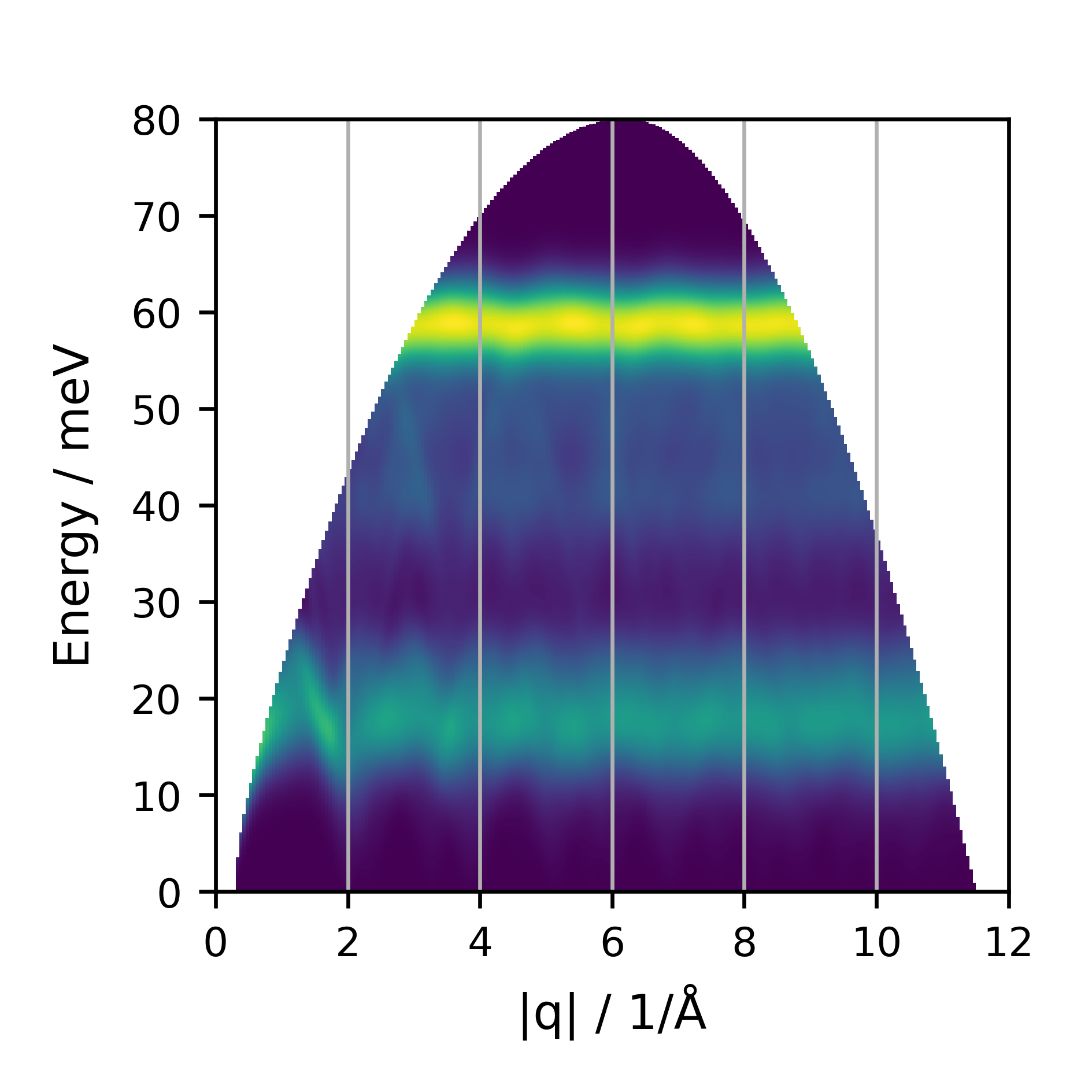}
        \caption{
        Fixed-width broadening: \SI{6}{\milli \electronvolt}
        \label{fig:merlin-6}
        }
    \end{subfigure}
    \\
        \begin{subfigure}[b]{\columnwidth}
        \includegraphics[width=0.9\columnwidth,trim={0.25cm 0.25cm 0.25cm 0.25cm},clip]{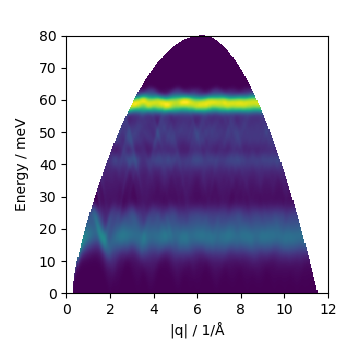}
        \caption{
        Variable-width broadening: Exact summation
        \label{fig:merlin-exact}}
    \end{subfigure}
            \begin{subfigure}[b]{\columnwidth}
        \includegraphics[width=0.9\columnwidth,trim={0.25cm 0.25cm 0.25cm 0.25cm},clip]{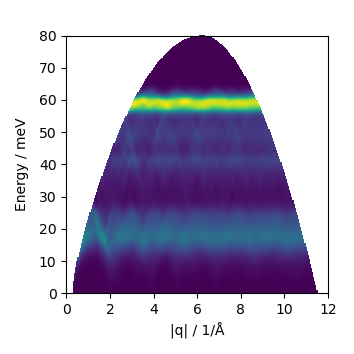}
        \caption{
        Variable-width broadening: Approximate
        \label{fig:merlin-approx}}
    \end{subfigure}
    \caption{Simulated INS of powdered Si on MERLIN instrument at ISIS.
    Powder-averaged coherent inelastic scattering function calculated with \codename{Euphonic}.
    Energy-dependent resolution function ranges \SIrange{1.93}{6.39}{\milli\electronvolt}, computed with \codename{PyChop2}.
    In upper row these limits are illustrated by fixed-width broadening (\ref{fig:merlin-2}, \ref{fig:merlin-6}),
    while the bottom row compares the exact summation of a Gaussian kernel at each energy bin (\ref{fig:merlin-exact}) with the approximate interpolated method (\ref{fig:merlin-approx}), using a relatively large nominal error of \SI{10}{\percent}.
    \label{fig:merlin}
    }
\end{figure*}

The effect of this is demonstrated in \cref{fig:merlin},
which considers a hypothetical INS experiment to measure powdered elemental silicon on the MERLIN time-of-flight spectrometer at ISIS.
The energy-dependent resolution function was modelled with the \codename{PyChop2} routine in \codename{Mantid} with some relevant instrument parameters: the ``G'' chopper package running at \SI{200}{\hertz} and incident energy \SI{80}{\milli\electronvolt}.\cite{noauthor_mantid_2013}
The coherent inelastic neutron scattering function of Si was calculated with numerical powder averaging using \codename{Euphonic} from force constants in the Kyoto phonon database using Phonopy from Materials Project structure data (entry mp-149-20180417).\cite{togoPhononDatabaseKyoto2015,togoFirstPrinciplesPhonon2015,jain_commentary_2013}
Broadening is applied with fixed-width Gaussian functions with FWHM \SI{2}{\milli\electronvolt} and \SI{6}{\milli\electronvolt}, and with the energy-dependent function ranging from \SIrange{1.93}{6.39}{\milli\electronvolt}.
With access to rapid simulation of resolution effects, it is possible to explore the impact of different instruments and measurement parameters before using expensive beamtime.

\section{Conclusions}
 A fast, approximate method is reported that alleviates the computational cost of variable-width broadening.
 This new approach significantly reduces the number of expensive Gaussian function evaluations, with an exact kernel calculated at logarithmically spaced intervals over the range of broadening widths.
 The input data is carefully distributed to small set of fixed-width broadening operations,
 such that a sum over the resulting spectra forms a good approximation to the broadening of each peak with its designated width.
 Due to interpolation errors and the pre-binning of data onto a regular grid, there is loss of accuracy compared to exact broadening and summation over individual peaks.
 However, with appropriate sampling density the error introduced is minimal, and even a large nominal error of \SI{20}{\percent} is shown to give visually acceptable results in some scientific applications.

The method is equally applicable to Lorentzian broadening and the method has been implemented for both Gaussian and Lorentzian broadening in the \codename{Euphonic} Python package, for accelerated \qpt{} sampling (by adaptive broadening) and simulation of instrumental resolution functions.
Further potential applications include interactive visualisation of resolution effects and the fitting of models to experimental data.

In its current form the method is only suitable for broadening along one axis at a time, but there is potential for the approach to be extended to higher dimensions.
Whereas in this 1-D scheme two basis functions are used at each point (i.e. a narrower and wider kernel of the same shape as the desired function),
a higher-dimensional scheme would likely benefit from the use of more general basis functions such as plane waves.
This would also support the implementation of more complex (e.g. bimodal) broadening functions.

\section{Data access statement}
Plots in this paper are generated with a set of Python scripts from simple synthetic data or with publicly-available force constant data from the Kyoto Phonon database.\cite{togoPhononDatabaseKyoto2015}
The scripts and pre-processed force-constant data files are available from the STFC Research Data Repository ``eData'' ({\tt https://edata.stfc.ac.uk/handle/edata/942}).
Some of these depend on the library implementation in \codename{Euphonic} v1.3.0 ({\tt DOI:0.5286/SOFTWARE/EUPHONIC/1.3.0}) which is available under the GNU General Public License v3 from Github and PyPI.

\section{Acknowledgements}
We acknowledge useful discussions with colleagues including Rebecca Fair (who also assisted with code review), Keith Refson, Duc Le, Sanghamitra Mukhopadhyay and Dominik Jochym.

\bibliographystyle{unsrt}
\bibliography{references, references2}

\end{document}


\maketitle

\section{Parametrisation of error/width relationship}
In order to aid selection of suitable kernel-width spacing, several error measures were calculated for a range of $\alpha$, geometrically spaced from 1.01 to 2.00.
These are obtained numerically by comparing a set of 20 linearly-spaced reference functions from $\sigma=1$ to $\sigma=\alpha$ to exact kernel evaluations on a 1001-point grid of $x$-values from -10 to 10.
(Strictly, $\sigma$ is only the width parameter for Gaussian kernels; for Lorentzian kernels the corresponding parameter was $\gamma$.)
As well as considering the average (integrated) and maximum (point-by-point) errors among the set of 20 function approximations, the largest integrated error of a single kernel in the set was taken and denoted as the "max integrated error".

\begin{figure}
    \centering
    \includegraphics[width=\columnwidth]{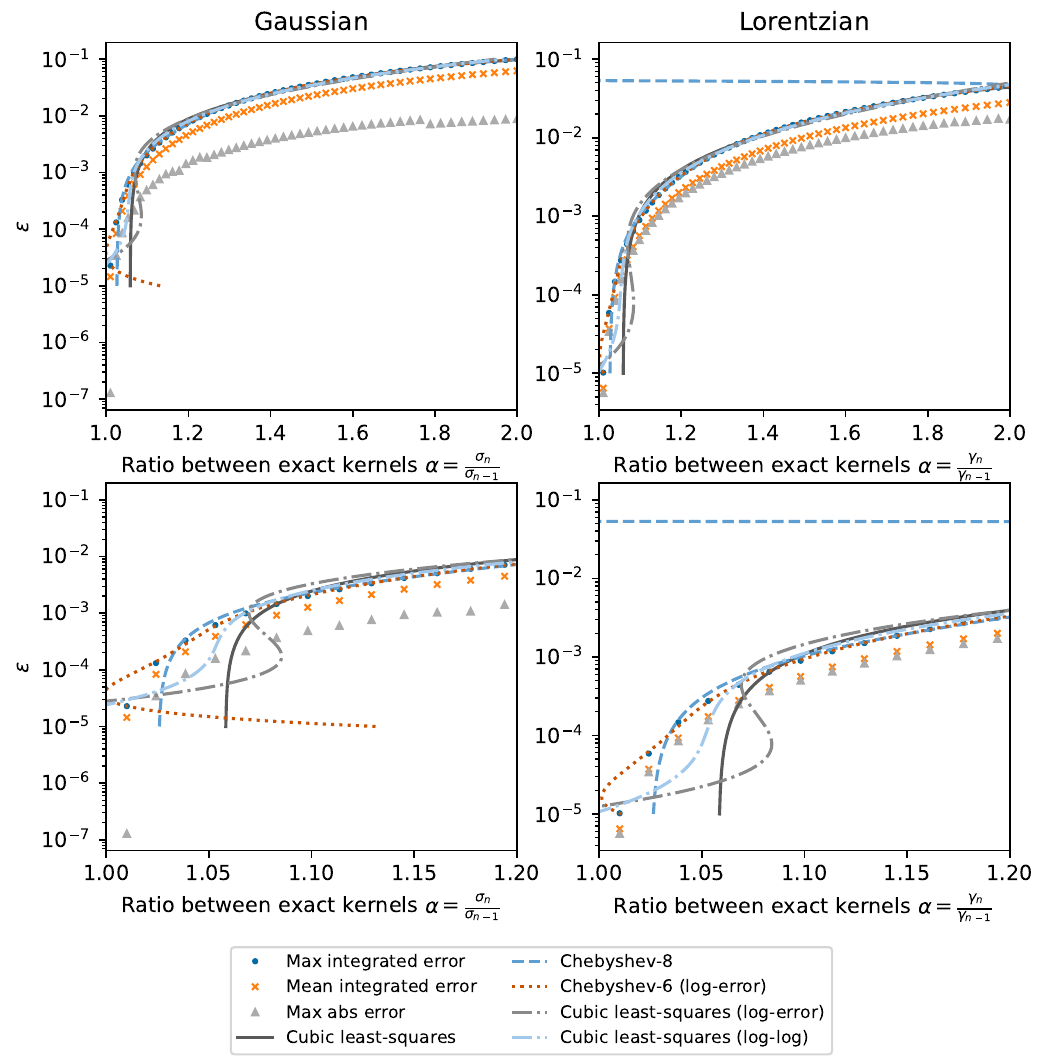}
    \caption{Relationship between kernel spacing and approximation errors, with a selection of polynomial fits to the "max integrated error". (i.e. the integrated error over the worst-performing interpolated kernel).}
    \label{fig:error-fits}
\end{figure}

These are plotted in \cref{fig:error-fits}, were we see that this "worst-case" error is also a smooth function and so a good target for parametrisation.
The choice of function form is arbitrary and a variety of fits are shown.
Fits to the log of the error tends perform effectively but with strange artefacts at the very smallest error values;
we also see that the order-8 Chebyshev (minimax) fit is unstable at the upper limit of the measured error values.
It is clearly important to constrain the application of these parametrisations to a well-behaved region.

A Python script is included with this SI which can be used to generate \cref{fig:error-fits} and report the corresponding fit parameters, or modified to vary the polynomial order and sampling density.

\begin{figure}
    \centering
    \includegraphics{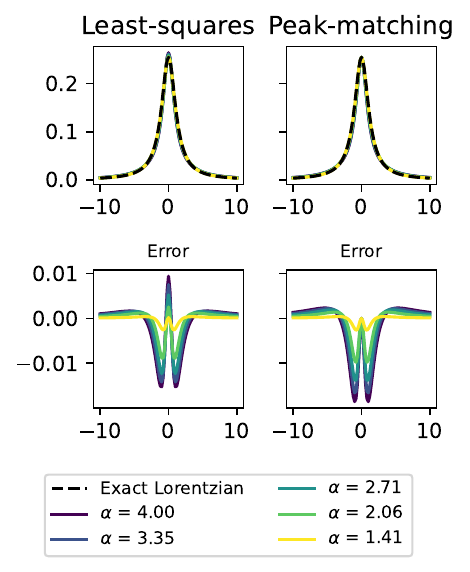}
    \caption{Variation in overall accuracy of Lorentzian approximation with width factor.
      Upper plots show approximations to Gaussian with $(\gamma / 2) =1.25$ produced using a linear combination of two Lorentzians with $(\gamma / 2)$ spaced by the specified width factor. Lower plots are obtained by subtracting exact function from approximations.
    The left column uses least-squares optimisation over whole data range to determine linear combination weights, whereas the on the right-hand-side the mixing parameters are chosen to match the exact peak height at $x = 0$.
    }
    \label{error_dist_lorentz}
\end{figure}